\documentclass[10pt,journal,twocolumn,twoside]{IEEEtran} 
\usepackage{xpatch}
\usepackage{graphicx}
\usepackage{subfigure}
\usepackage{epstopdf}
\usepackage{float}
\usepackage{algorithmic}
\usepackage{array}
\usepackage{amsmath}
\usepackage{amssymb}
\usepackage{mdwmath}
\usepackage[table]{xcolor}
\usepackage{makecell}
\usepackage{eqparbox}
\usepackage{stfloats}
\usepackage{fixltx2e}
\usepackage{tabularx}
\usepackage{cases} 
\usepackage{booktabs}

\usepackage{flushend}

\usepackage{threeparttable}
\usepackage{xcolor}
\usepackage{makecell}
\usepackage{multirow}

\usepackage[boxed,ruled,commentsnumbered]{algorithm2e}
\usepackage{url}
\usepackage{cite}
\allowdisplaybreaks[4]

\makeatletter

\newcommand{\Rmnum}[1]{\expandafter\@slowromancap\romannumeral #1@}
\makeatother

\makeatletter
\renewcommand*{\@opargbegintheorem}[3]{\trivlist
      \item[\hskip \labelsep{\bfseries #1\ #2}] \textbf{(#3):}\ }
\makeatother

\begin{document}

\makeatletter
\def\changeBibColor#1{%
  \in@{#1}{}
  \ifin@\color{red}\else\normalcolor\fi
}
 
\xpatchcmd\@bibitem
  {\item}
  {\changeBibColor{#1}\item}
  {}{\fail}
 
\xpatchcmd\@lbibitem
  {\item}
  {\changeBibColor{#2}\item}
  {}{\fail}
\makeatother

\title
{Semantic Importance-Aware Communications with Semantic Correction Using Large Language Models
}
\author{Shuaishuai Guo,~\IEEEmembership{Senior Member, IEEE}, Yanhu Wang, Jia Ye,~\IEEEmembership{Member IEEE}, Anbang Zhang, \\Peng Zhang, and Kun Xu
\thanks{S. Guo, Y. Wang and A. Zhang are with the School of Control Science and Engineering, and also with Shandong Key Laboratory of Wireless Communication Technologies, Shandong University, China (e-mail: shuaishuai$\_$guo@sdu.edu.cn; yh-wang@mail.sdu.edu.cn; 202234946@mail.sdu.edu.cn). }
\thanks{Jia Ye is with the School of Electrical Engineering, Chongqing University, Chongqing, 400044, China (jia.ye@cqu.edu.cn).}
\thanks{Peng Zhang is with the School of Physics and Electronic Information, Weifang University, Weifang, 261061, China (e-mail: sduzhangp@163.com).
}
\thanks{Kun Xu is with the Department of Government and Enterprise, China Mobile Communications Group, Shandong Co. LTD, Jinan 250001, Shandong, China (e-mail: sduzhangp@163.com).
}
\thanks{The code can be available at \protect\url{https://github.com/SSG2019/ULSC}.}
   }
\maketitle


\begin{abstract} 
Semantic communications, a promising approach for agent-human and agent-agent interactions, typically operate at a feature level, lacking true semantic understanding. This paper explores understanding-level semantic communications (ULSC), transforming visual data into human-intelligible semantic content. We employ an image caption neural network (ICNN) to derive semantic representations from visual data, expressed as natural language descriptions. These are further refined using a pre-trained large language model (LLM) for importance quantification and semantic error correction. The subsequent semantic importance-aware communications (SIAC) aim to minimize semantic loss while respecting transmission delay constraints, exemplified through adaptive modulation and coding strategies. At the receiving end, LLM-based semantic error correction is utilized. If visual data recreation is desired, a pre-trained generative artificial intelligence (AI) model can regenerate it using the corrected descriptions. We assess semantic similarities between transmitted and recovered content, demonstrating ULSC's superior ability to convey semantic understanding compared to feature-level semantic communications (FLSC). ULSC's conversion of visual data to natural language facilitates various cognitive tasks, leveraging human knowledge bases. Additionally, this method enhances privacy, as neither original data nor features are directly transmitted.

\end{abstract}


\section{Introduction} 
\IEEEPARstart{R}{ecent} developments in semantic communications, previously underexplored or overlooked, have propelled this field to the forefront of scholarly inquiry\cite{GuoVTC}.
 This renewed focus is driven by several key factors: advancements in deep learning technologies, the evolution of applications involving both agent-human and agent-agent communications, and a growing recognition of its critical role in transcending the limitations of system designs governed by Shannon's communication theory\cite{strinati20216g,lan2021semantic,tong2022nine,9955525}. 

\subsection{Related Work}
The evolution of semantic communications has been marked by a transition from transmitting original data to channel-robust, low-dimensional latent space representations. This paper delineates three primary categories based on the nature of these representations.
\subsubsection{Continuous Latent Space Representations}
Continuous latent space representations involve encoding original data into continuous-valued vectors. Various methodologies, including autoencoder (AE)\cite{zhai2018autoencoder}, variational autoencoder (VAE)\cite{kingma2013auto}, and generative adversarial network (GAN)\cite{2014Generative}, facilitate this process.
These methods enable the encoding of source data into a compact continuous space, emphasizing essential features while omitting redundant elements, thereby optimizing transmission efficiency. This approach aids in tasks such as retrieval, classification\cite{10458014}, and reconstruction \cite{9953099},
, as exemplified in studies focused on image reconstruction \cite{bourtsoulatze2019deep}  and image retrieval \cite{Jankowski2021}, which demonstrate enhanced performance compared to traditional communication systems. Recently, semantic communication applications have expanded to encompass text \cite{Xie2021,Jiang2022Aug}, audio \cite{Weng2021Aug, 9953316}, and video \cite{jiang2022wireless}, showcasing promising results. However, integration with existing digital communication systems remains a challenge due to compatibility issues, necessitating substantial system-wide updates \cite{GuoS2023}.

\subsubsection{Discrete Latent Space Representations}
This category involves transforming continuous semantic vectors into discrete codewords, aligning with digital communication systems' structures. Xie \emph{et al.} \cite{xie2020lite} developed quantization-based feasible constellations for existing systems. The VAE's Gaussian latent variable distribution was modified to a Bernoulli distribution for discrete representation acquisition \cite{2018NECST}, although this added complexity to the loss computation. Nemati \emph{et al.} \cite{nemati2023vq} introduced a vector quantized VAE (VQ-VAE) based joint source and channel coding approach, employing a shared codebook between transmitter and receiver for mapping semantic vectors to codewords. This method effectively clusters semantic vectors, with codewords serving as class centers. Subsequent studies \cite{fu2023vector, xie2023robust,10101778,10483054} have proposed variations in encoder and decoder architectures, while retaining the discrete representation principle through codebook construction.

\subsubsection{Natural Language Representations}
Contrasting with feature-level systems, this approach addresses the semantic gap between low-level data and high-level human-associated meanings. It involves mapping data to human-comprehensible semantic domains using natural language representations. Despite their redundancy, these representations offer numerous benefits. They align with human knowledge bases, enabling semantic encoding and decoding that reliably convey meanings \cite{GuoS2023}. This paper further explores semantic-aware adaptive modulation and coding, employing pre-trained language models for semantic error correction at the receiver. Converting visual data into natural language facilitates tasks like commonsense reasoning and decision-making at both transceiver ends. It also enhances applications involving generative AI or human receivers, such as aiding visually impaired individuals or improving autonomous systems' interaction with visual data. Furthermore, this method offers privacy advantages, as neither original data nor features are directly transmitted, complicating unauthorized inference of the data's semantics.


\subsection{Contributions of This Work}
This paper introduces a novel understanding-level semantic communications (ULSC) system, adeptly designed for seamless integration with existing communication networks. The system is characterized by its utilization of advanced components, including pre-trained image caption neural networks (ICNNs), natural language descriptions, and language models, alongside an adaptive communication strategy and semantic error correction at the receiver. The salient contributions of this work are outlined as follows:

\begin{itemize}
\item The deployment of a pre-trained ICNN facilitates the conversion of visual data into textual representations in natural languages, significantly reducing bandwidth requirements and enhancing accessibility for individuals with visual impairments. These textual descriptions also offer enhanced searchability and indexation capabilities for large datasets and serve as a means to safeguard privacy by transmitting information devoid of sensitive visual details.

\item Leveraging a pre-trained language model, specifically BERT (Bidirectional Encoder Representations from Transformers), we quantify the semantic importance of frames within textual descriptions. This process underpins an adaptive communication strategy, prioritizing the transmission of essential information. The system conducts frame-level, semantic importance-aware communications, encompassing data compression, channel coding, and various network operations. This approach is exemplified through the investigation of adaptive modulation and coding, with a focus on performance evaluation and system optimization in frame erasure channels.

\item At the receiver end, we propose the use of a pre-trained language model, trained on mask modeling, for the completion of erasure-impacted frames. Utilizing BERT's mask language modeling, we address missing frame segments by predicting and replacing words, thereby enhancing semantic recovery capabilities. This method underscores the benefits of employing natural language representations in ULSC, adding an additional layer of error correction.


\item We conduct a thorough evaluation of semantic similarities between transmitted and recovered semantic understanding, comparing our ULSC system with traditional feature-level digital semantic communication systems. The paper also assesses the inference delay of pre-trained AI models utilized in ULSC and demonstrates the system's added benefit in terms of privacy preservation.
\end{itemize}

\begin{figure*}
       \centering       \includegraphics[width=1\linewidth]{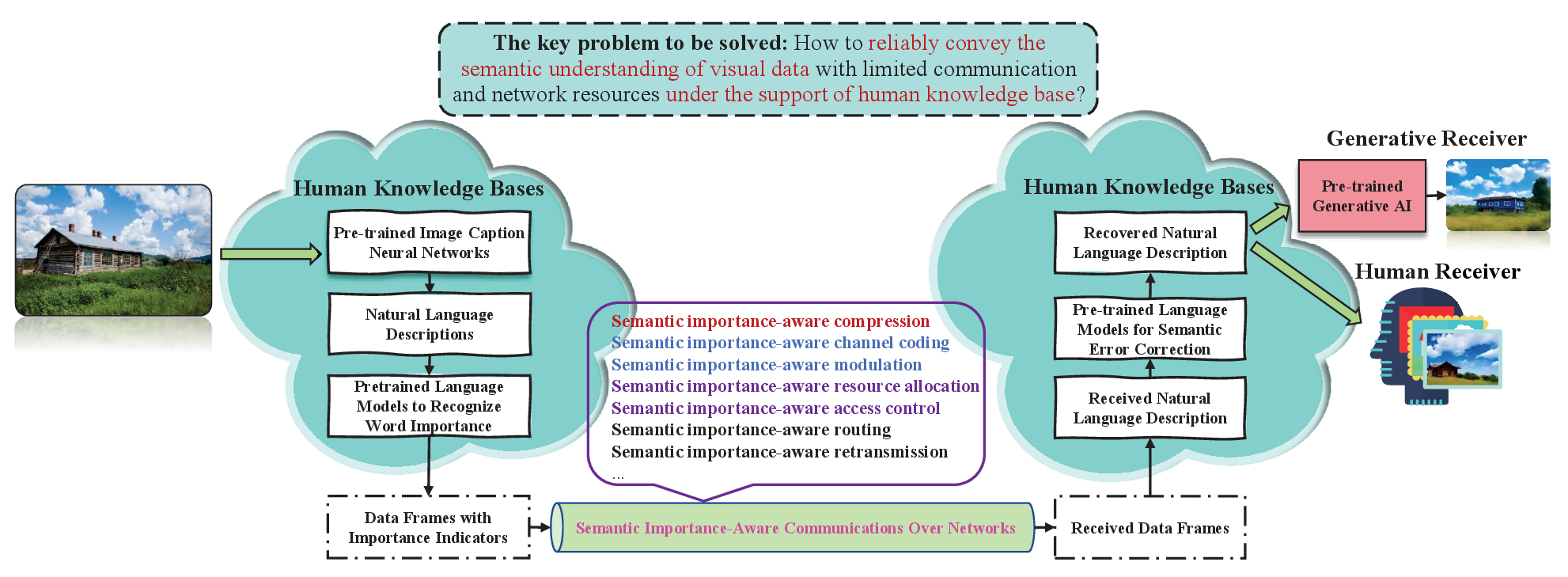}       \caption{
Understanding-level semantic communications to reliably transmit the semantic understanding of visual data with limited communication between the transmitter and the receiver under the support of human knowledge bases.}
       \label{ULSC}
\end{figure*}

\subsection{Organization}

The remainder of this paper is methodically structured to guide readers through our comprehensive research in ULSC. Section II begins with an in-depth examination of the ULSC system model, detailing the transceiver structures and mathematical modeling essential to our approach. This is followed by Section III, which delves into the transmitter functions, particularly focusing on the conversion of visual data to natural language and the quantification of frame importance. Section IV discusses the implementation of frame-level, semantic importance-aware communications over networks, incorporating a thorough performance evaluation and system optimization in the context of frame erasure channels. In Section V, the paper shifts focus to the receiver's role, elaborating on the use of a pre-trained language model for semantic error correction and the application of pre-trained generative AI models for visual data generation. Section VI presents our experimental studies and discussions, offering crucial insights into the practical applications and effectiveness of the proposed system.Finally, the paper culminates in Section VII with conclusions that summarize our key findings and contributions, while also providing a perspective on potential future directions in this evolving field of research.

\section{System Model}
In this paper, we consider a ULSC system as illustrated in Fig. \ref{ULSC}. This system encompasses a sophisticated agent, equipped with visual input capabilities, functioning as a transmitter. This transmitter engages in communication either with an individual who is visually impaired or with a generative artificial intelligence model. The primary objective of the ULSC framework is to facilitate the transmission of the semantic essence of visual information, achieving this aim within the constraints of limited communication bandwidth and network resources, and augmented by the integration of human knowledge bases.
To accomplish this, the visual information at the transmitter is initially processed through an ICNN. This processing translates the visual data into coherent natural language representations. This approach enables the encapsulation of complex visual content into linguistically accessible formats, thereby bridging the gap between visual perception and semantic communication.

Let $\mathbf{S}\in\mathbb{R}^N$ denote the $N$-dimensional array representing the input visual data, and let $\mathbf{X}\in\mathbb{Z}^D$ signify the output, which is a 
 $D$-word natural language representation of the visual data. In this context, $\mathbb{Z}$ is identified as the vocabulary set. The transformation from the input visual data to its natural language representation is governed by the following equation:
\begin{equation}
\mathbf{X}={f}_{\alpha}(\mathbf{S}),
\end{equation}
where $f_{\alpha}(\cdot)$ symbolizes the ICNN, and $\alpha$ the parameters of the ICNN.
For transmission, the $D$ words of the natural language representation are partitioned into 
 $L$ distinct groups. Each group is then encoded—this process encompasses both source and channel coding—into $L$ separate binary sequences, denoted as $\mathbf{b}_1,\mathbf{b}_2,\cdots, \mathbf{b}_L$. which will the payload of $L$ frames to be transmitted. These binary sequences form the payload of 
$L$ individual frames, which are then prepared for transmission. This segmentation and encoding strategy is pivotal in optimizing the transmission process, ensuring both efficiency and reliability in the conveyance of the semantic information encapsulated within the visual data.

In this research, we introduce the concept of semantic importance, denoted as $F_l$
 , for the $l$th frame in the context of ULSC. This parameter, 
$F_l$, plays a critical role in guiding the comprehensive design of each frame. The design elements influenced by the frame importance 
$F_l$ include the payload bit sequence $b_l$  (encompassing source and channel coding strategies), the application header (incorporating error concealment mechanisms), the transport header (addressing re-transmission protocols), the Internet Protocol (IP) header (focusing on routing configurations), and the link header (which includes aspects like access control, resource allocation, and modulation strategies).

The framework of semantic importance-aware communications diverges significantly from traditional communication methods. Its distinctiveness lies in the incorporation of semantic importance in the communication process. This approach allows for the selection of diverse communication technology combinations, specifically tailored to ensure higher reliability in the transmission of semantically important frames. In these designs, there is a dual focus: one on the uncertainty of bits as carriers of information and the other on the semantic significance underlying these bits.

A notable aspect of the proposed designs is their compatibility with existing communication networks. This compatibility signifies that ULSC communications can be seamlessly integrated and operated over current network infrastructures, thereby enabling a smooth and gradual evolution of communication systems. This adaptability ensures that the advanced capabilities of ULSC, particularly in terms of semantic importance, can be leveraged without necessitating a complete overhaul of existing communication frameworks.

Upon traversing the resource-constrained network channels, the frames reach the receiver, where they undergo a sequence of processing steps. These steps include demodulation, channel decoding, and source decoding, which are integral for the extraction of the natural language representations from the transmitted frames. It is important to note that during this transmission process, the natural language representations may become corrupted or, in some cases, portions of them may be lost.

To address these potential discrepancies and ensure the integrity of the transmitted information, we propose the integration of a large language model (LLM). This LLM is pre-trained on a comprehensive human textual knowledge base, thereby equipping it with robust capabilities for semantic error correction. The LLM's advanced algorithms enable it to effectively identify and rectify semantic errors or gaps in the received natural language representations, thereby ensuring that the conveyed information remains coherent and contextually accurate. This process can be formalized as:
\begin{equation}\label{SemCorrection}
\hat{\mathbf{X}} = f_{\theta}(\Tilde{\mathbf{X}}),
\end{equation}
where $\Tilde{\mathbf{X}}$ and $\hat{\mathbf{X}}$ are the corrupted natural language representations and corrected natural language representations respectively. $f_{\theta}(\cdot)$ denotes the LLM designated for semantic correction, with $\theta$ representing its parameters.

Following the successful recovery and correction of the natural language representations, these representations are then conveyed to the end recipient. The recipient could be a person with visual impairment, to whom the information is transmitted via an acoustic channel. Alternatively, the recipient could be a generative AI model, where the natural language representations serve as textual prompts.
This process can be formalized as:
\begin{equation}\label{VDG}
\hat{\mathbf{S}} = f_{\beta}(\hat{\mathbf{X}}),
\end{equation}
where $\hat{\mathbf{S}}$ represents the generated speech, image, or other formats of content. $f_{\beta}(\cdot)$ denotes speech generation model, generative AI model, or some other model, and $\beta$ represents its parameters. This flexibility in the mode of information delivery underscores the versatility of the system, making it adept at catering to varied user needs and scenarios. The use of an acoustic channel for visually impaired individuals is particularly noteworthy, as it transforms visual data into an accessible auditory format, thereby enhancing information accessibility for users with visual limitations.


\begin{figure}[t]
\centering
\vspace{-0.35cm}
\subfigtopskip=2pt
\subfigbottomskip=2pt
\subfigcapskip=-2pt
\subfigure[The pre-trained ICNN.]{
\centering \includegraphics[width=1\linewidth, keepaspectratio=false]{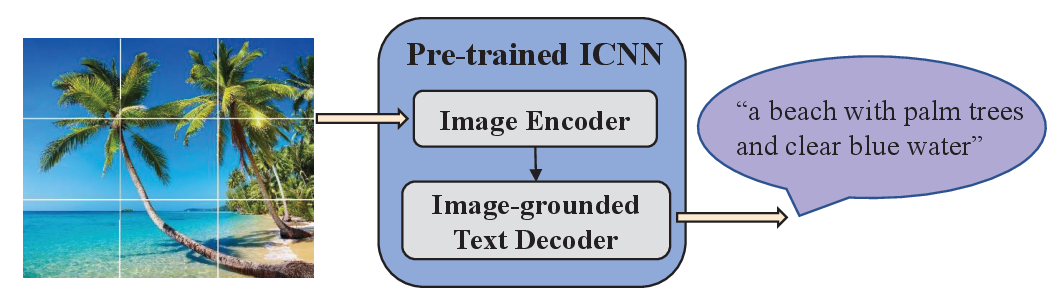}
}
\vspace{0.25cm}
\subfigure[Image Encoder and Image-grounded Text Decoder.]{
\centering \includegraphics[width=1\linewidth, keepaspectratio=false]{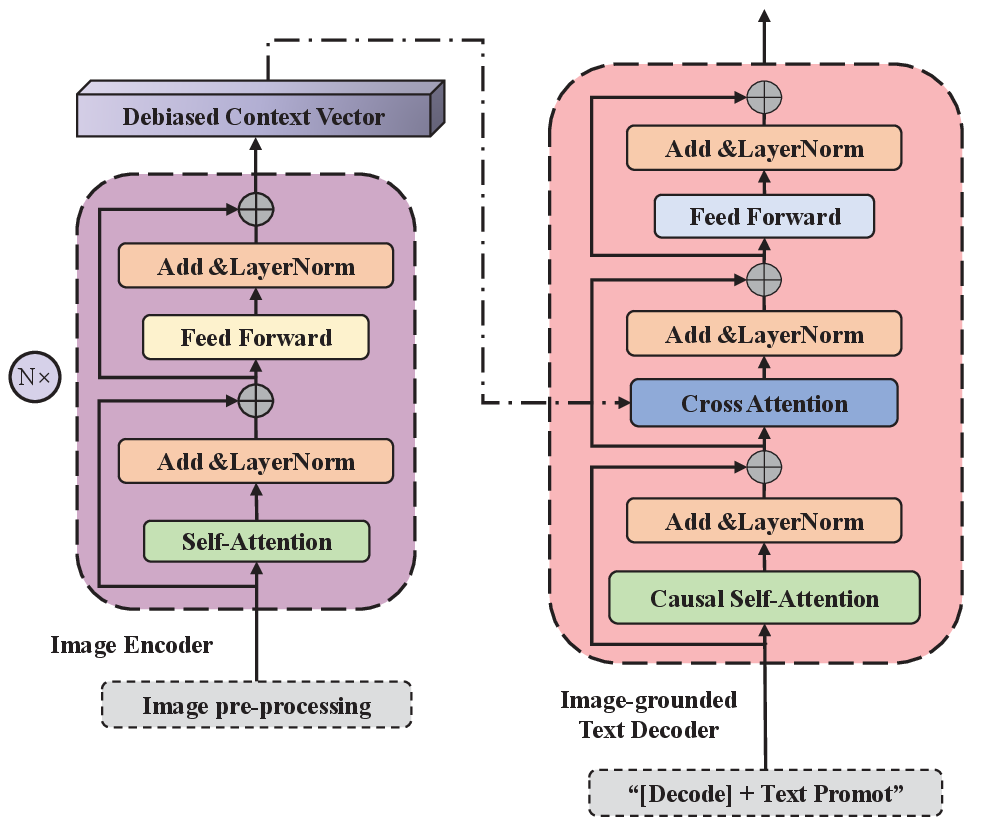}
}
\caption{The adopted pre-trained ICNN and its details.}
\label{three graphs}
\end{figure}

\section{Textual Encoder and Frame Importance Quantification}
This section delves into the intricacies of the textual encoder within the visual data-to-text conversion process and the methodology for quantifying frame importance, which are fundamental components of the ULSC transmitters.

\subsection{Visual Data to Natural Language Conversion}
The primary objective of ULSC is to facilitate the comprehension of visual data for visually impaired individuals or generative AI systems by utilizing expansive human knowledge bases. The transformation of visual data into natural language is a critical and intricate process within ULSC, encompassing aspects of computer vision (CV) and natural language processing (NLP). This complex process, referred to as image captioning in the field of deep learning, requires a deep understanding of visual content to generate accurate and meaningful descriptions.

Significant advancements have been made in addressing this challenge. Notable approaches include the encoder-decoder model, as demonstrated in works by Vinyals et al. \cite{vinyals2016show} and Karpathy et al. \cite{Karpathy2017D}. These models employ convolutional neural networks (CNNs) to extract image features and recurrent neural networks (RNNs) or long short-term memory (LSTM) networks to generate corresponding textual descriptions. To enhance the accuracy and relevance of these descriptions, an attention mechanism has been integrated into the encoder-decoder framework. This mechanism allows the model to focus on different segments of the image during the generation process, as detailed in the studies by Xu et al. \cite{xu2015show} and Li et al. \cite{li2022blip}.

Building on these advancements, our research has adopted the pre-trained ICNN proposed by Li et al. \cite{li2022blip}. Fig. \ref{three graphs}(a) shows the workflow of the ICNN. The input image is divided into a number of small blocks, which are converted into a high-dimensional feature representation after passing through the image encoder. These image feature vectors are then passed to an image-based text decoder, which generates a text description related to the image content. Fig. \ref{three graphs}(b) shows the structure of the image encoder and image-grounded text decoder in detail. The image encoder consists of $N$ Transformer layers. Each Transformer layer includes a Self-Attention layer, Feed Forward network, and LayerNorm. These layers extract image features by Self-Attention mechanism, and further process them by Feed Forward network, and finally generate image feature vectors. In image-grounded text decoder, each Transformer layer also contains LayerNorm, Feed Forward network, and Causal Self-Attention layer. In addition, a Cross Attention layer is added, which is used to combine image features and text information to generate text descriptions related to the image content. Further details are available in \cite{li2022blip}.

\subsection{Frame Importance Quantification}

The concept of frame importance is pivotal in semantic importance-aware communications. It pertains to the relevance and significance of data within a frame for the comprehensive understanding of the overall context. Due to its content-dependent and dynamic nature, the quantification of semantic importance is intricate and often requires deep contextual understanding and domain-specific knowledge. Establishing universal metrics for this quantification is challenging, as it may vary across different contexts.

In our approach, we amalgamate human judgment with automated methods, including natural language processing techniques, machine learning algorithms, and user feedback, to quantify the semantic importance of data within frames. This multifaceted approach is a subject of ongoing research. Specifically, we utilize pre-trained language models, such as ChatGPT and BERT, to measure the importance of words and frames based on the semantic variations resulting from their erasure, as detailed in \cite{GuoS2023}.

Our method involves comparing the original message, denoted as $\mathbf{m}_{0}$, with a modified version, $\mathbf{m}_{l}$, where the $l$th frame is absent but replaced with a completion frame generated by a LLM. The semantic importance of the $l$th frame is then quantified as:
\begin{equation}\label{eq2}
w_l = 1 - \phi(\mathbf{m}_{0}, \mathbf{m}_{l}),
\end{equation}
where $\phi\left(\mathbf{m}_{0}, \mathbf{m}_{l}\right) =\frac{B_{\boldsymbol{\psi}}\left(\mathbf{m}_{0}\right)^T B_{\boldsymbol{\psi}}\left(\mathbf{m}_{l}\right)}{\left\|B_{\boldsymbol{\psi}}\left(\mathbf{m}_{0}\right)\right\| \cdot\left\|B_{\boldsymbol{\psi}}\left(\mathbf{m}_{l}\right)\right\|}$ measures the semantic similarity between $\mathbf{m}_{0}$ and $\mathbf{m}_{l}$. This similarity is calculated using the cosine similarity of the embeddings generated by a pre-trained BERT model, $B_{\boldsymbol{\psi}}(\cdot)$,  for both messages. This approach ensures that the quantification of importance reflects the ability of semantic error correction mechanisms to recover or predict missing information.

Notably, this method allows the transmitter to independently assess the importance of words or frames without requiring interaction with the receiver. By simulating a semantic error corrector, the transmitter can determine the potential for each word or frame to be accurately reconstructed or predicted by the receiver's error correction system.

Table \ref{T1} presents a comparative analysis between our proposed method, which includes semantic error correction, and the previously discussed method that does not incorporate this aspect (as detailed in \cite{GuoS2023}). A striking observation from this comparison is the prevalence of zero importance scores for most words in the context of semantic error correction. This outcome indicates that many words, when lost, can be effectively predicted and completed by the receiver's error correction mechanisms. Consequently, the ULSC framework can prioritize transmitting words or frames that are less likely to be accurately reconstructed if lost, thereby enhancing communication efficiency and reliability.

This approach represents a significant advancement in the field of semantic communications, particularly in applications involving ULSC, as it provides a more nuanced understanding of data importance and optimizes data transmission based on the receiver's error correction capabilities.

\begin{table*}[t]
\footnotesize
\caption{The Importance of Each Word in Example Sentences With and Without Semantic Error Correction.}
\label{T1}
\centering
\renewcommand{\arraystretch}{1.5}
\begin{tabular}{|c|l|c|c|c|c|c|c|c|c|c|}
\hline
\multirow{3}{*}{\textbf{Example 1}} &\textbf{Word} & A &beach &with &palm &trees &and &clear &blue &water\\
\cline{2-11}
&\textbf{Semantic Error Correction} & 0 & 0.0386 & 0 & 0& 0 & 0 & 0 & 0 & 0.0457 \\
\cline{2-11}
&\textbf{No Semantic Error Correction \cite{GuoS2023}} & 0.0197  & 0.0515 & 0.0389 & 0.0131 & 0.0253 & 0.0357 & 0.0109 & 0.0145 & 0.0342\\
\hline
\multirow{3}{*}{\textbf{Example 2}}&\textbf{Word} &A  &bald  &eagle  &flying  &over  &a  &body  &of  &water\\
\cline{2-11}
&\textbf{Semantic Error Correction} & 0 & 0.0092 & 0.0780 & 0.0593& 0 & 0 & 0 & 0 & 0.1045\\
\cline{2-11}
&\textbf{No Semantic Error Correction \cite{GuoS2023}} & 0.0227  & 0.0130 & 0.0758 &  0.0474  & 0.0229  &0.0438  & 0.0606 & 0.0399  & 0.0844\\
\hline
\multirow{3}{*}{\textbf{Example 3}}&\textbf{Word} & A &forest &filled &with &lots &of &tall &trees &---\\
\cline{2-11}
&\textbf{Semantic Error Correction} & 0 & 0.0364 & 0.0305 & 0& 0.0432 & 0 & 0.0104 & 0.0740 & ---\\
\cline{2-11}
&\textbf{No Semantic Error Correction\cite{GuoS2023}} & 0.0426  & 0.0423 & 0.0145 & 0.0244 & 0.0574 & 0.0219 & 0.0132 & 0.0477 & ---\\
\hline
\end{tabular}
\end{table*}

\section{Semantic Importance Aware Communications Over Networks}
The concept of semantic importance-aware communications centers around the prioritization of semantically significant frames during transmission. This approach is instrumental in optimizing data traffic management and ensuring that vital information is accorded the utmost priority. Such prioritization is crucial in minimizing delays and augmenting the overall efficiency of the system. In this framework, frames of greater semantic importance are transmitted or processed in precedence to those of lesser significance.

Semantic importance-aware communications exhibit versatility across multiple layers. At the foundational physical layer, diverse modulation and coding schemes are employed to transmit data of varying levels of importance. For example, data of higher importance might be encoded using more robust error-correcting codes or transmitted using lower-order modulation to enhance reliability.

In specific physical layer technologies, such as cellular systems, there exists the capability for dynamic allocation of channels to frames based on their importance. Frames of higher importance may be allocated to channels experiencing less congestion. This strategy, situated at the physical layer, encompasses the retransmission of data in instances of errors or collisions during the initial transmission. In scenarios where significant data encounters errors, it may be retransmitted with elevated priority to assure a more reliable delivery.

Furthermore, power control mechanisms within wireless communication can allocate preferential treatment to certain frames by modulating their transmit power. This adjustment ensures that frames of higher importance receive better signal quality and encounter reduced interference.

It is imperative to recognize that while some of these methodologies are implemented directly within the physical layer, many communication systems integrate both physical layer and higher-layer techniques, such as importance-aware routing and access control, to realize efficient and reliable semantic communication.

To elucidate semantic importance-aware communication and its integration with existing communication frameworks, we examine the semantic importance-aware adaptive modulation and coding selection (MCS) as a representative example.  In this model, transmitted messages are divided into a total of $L$ frames, each capable of carrying a payload of $A_l$ bits, inclusive of both header and data ($\mathbf{b}_l$). This semantic communication system deviates from traditional systems that predominantly focus on the accuracy of transmitted bits or symbols. Instead, it emphasizes the paramount objective of enabling the receiver to accurately comprehend and extract the intended information relayed by the transmitter. In this context, traditional metrics such as bit/symbol/frame error or loss probability are insufficient as evaluation criteria. Instead, we advocate for the adoption of semantic loss as a metric to quantify the integrity of the information exchange between the transmitter and receiver.

The expected semantic loss ($SL$)  incurred by a frame loss is determined by the semantic importance of the messages contained within each frame ($w_l$), combined with the frame loss probabilities ($P_{l, frame}^{Loss}$).  
It is essential to acknowledge that, in the sphere of semantic communications, each frame conveys unique messages, imparting varying degrees of semantic significance. This variation in importance amongst frames, often overlooked in conventional data-oriented communication systems, is of paramount importance in semantic communications. Notably, the omission or inaccuracy of messages bearing critical information can lead to a disproportionately high semantic loss. Therefore, the expected semantic loss due to frame loss is characterized by the integration of the semantic importance of the messages and their corresponding loss probabilities as
\begin{equation}\label{eq1}
SL=\sum_{l=1}^{L} w_l P_{l,frame}^{Loss}.
\end{equation}

The frame loss probability hinges upon the MCS and the channel status. Let $P_{l,frame}^{Correct}$ denote the successful reception and recovery of the $l$th frame and we have
\begin{equation}
P_{l,frame}^{Loss}=1-P_{l,frame}^{Correct}.
\end{equation}
The correct probability can be derived as
\begin{equation}
P_{l,frame}^{Correct}=(P_{l,bolck}^{Correct})^{N_B}.
\end{equation}
where $P_{l,bolck}^{Correct}$ is the correct probability of coding block. $N_B=\lceil{A_l}/{(R_lB_l)}\rceil$ is the number of coding blocks per frame, where $R_l$ is the coding rate, $B_l$ is the coding block length.
\begin{equation}\label{Plframe}
\begin{aligned}
&P_{{l,block}}^{Correct}=\\&\sum_{i=0}^{C_l}\left(B_l\atop i\right) P_{l,bit}^{Error}\left(\mathcal{M}_l,\gamma\right)^i\left(1-P_{l,bit}^{Error}\left(\mathcal{M}_l,\gamma\right)\right)^{B_l-i},
\end{aligned}
\end{equation}
where $C_l$ denotes the number of corrigible errors and it is determined by the choosing coding scheme. In general, $C_l = \lfloor\frac{d_{l, \min}-1}{2}\rfloor$, where $d_{l, \min}$ represents the minimum number of bit positions by which any two distinct codewords differ. $P_{l,bit}^{Error}\left(\mathcal{M}_l,\gamma\right)$ is the bit error probability determined by the symbol error probability $P_{l,sym}^{Error}\left(\mathcal{M}_l,\gamma\right)$, and the adopted bit-to-symbol mapping scheme. It can be approximated accurately as $P_{l,bit}^{Error}\left(\mathcal{M}_l,\gamma\right) \simeq \frac{P_{l,sym}^{Error}\left(\mathcal{M}_l,\gamma\right)}{\log_2 M_l}$ in high SNR region in case of Gray mapping, where $M_l$ denotes the modulation order for the adopted $\mathcal{M}_l$ modulation scheme. $\gamma=|h|^2 E_bR_l\log{M_l}/N_0$ represents the SNR, where $h$ is the channel coefficient, $E_b$ stands for power per bit, and $N_0$ denotes the noise power.
The  SEP of typical modulation schemes has been summarized in Table \ref{CSEP}. 
\begin{table*}[!t]
\centering
\renewcommand{\arraystretch}{2}
\caption{SEP of Various Modulation Schemes.}
\setlength{\tabcolsep}{10mm}{
\begin{tabular}{ |c|c| }
\hline Schemes&  $P_{l,sym}\left(\mathcal{M}_l, \gamma\right)$\\
 \hline
  $M$-PSK & $Q\left(\sqrt{2\gamma }\right)+\frac{2}{\sqrt{\pi}} \int_0^{\infty} \exp \left[-\left(u-\sqrt{\gamma}\right)^2\right] Q\left(\sqrt{2} u \tan \frac{\pi}{M}\right) d u $\\
  \hline 
  $M$-QAM & $4\left(\frac{\sqrt{M_l}-1}{\sqrt{M_l}}\right) Q\left(\sqrt{\frac{3 \gamma}{(M_l-1)}}\right)-4\left(\frac{\sqrt{M_l}-1}{\sqrt{M_l}}\right)^2 Q^2\left(\sqrt{\frac{3 \gamma}{(M_l-1)}}\right)$\\
 \hline
 
\end{tabular}}
\label{CSEP}
\end{table*}

Therefore, based on the derived symbol error probability, we can obtain the expression of the frame loss probability  as
\begin{equation}\label{eq7}
\begin{split}
P_{l,frame}^{Loss} 
 =1-&\left[\sum_{i=0}^{C_l}\left(A_l\atop i\right) \left(\frac{P_{l,sym}\left(\mathcal{M}_l,\gamma\right)}{\log_2 M_l}\right)^i\right.\\
 &~~~~\times\left.\left(1-\frac{P_{l,sym}\left(\mathcal{M}_l,\gamma\right)}{\log_2 M_l}\right)^{A_l-i}\right]^{N_B}.
\end{split}
\end{equation}

Substituting (\ref{eq2}) and (\ref{eq7}) into (\ref{eq1}) yields
\begin{equation}
\begin{split}
&SL
=\\&\sum_{l=1}^{L}\left(1-\frac{B_{\boldsymbol{\psi}}\left(\mathbf{m}_{l,t}\right)^T B_{\boldsymbol{\psi}}\left(\mathbf{m}_{l,r}\right)}{\left\|B_{\boldsymbol{\psi}}\left(\mathbf{m}_{l,t}\right)\right\| \cdot\left\|B_{\boldsymbol{\psi}}\left(\mathbf{m}_{l,r}\right)\right\|}\right)\left\{1-\left[\sum_{i=0}^{C_l}\left(A_l\atop i\right) \right.\right.
\\
&~~\times\left.\left.\left(\frac{P_{l,sym}\left(\mathcal{M}_l,\gamma\right)}{\log_2 M_l}\right)^i\left(1-\frac{P_{l,sym}\left(\mathcal{M}_l,\gamma\right)}{\log_2 M_l}\right)^{A_l-i}\right]^{N_B}\right\}.  
\end{split}
\end{equation}

To ensure that the semantic communication system meets its latency requirements, it is imperative to assess the overall transmission time for the conveyed messages. Assuming a transmission rate of $R_s$ in symbols/seconds for the semantic system under consideration, the transmission time for the $l$-th frame can be determined as
\begin{equation}
T_l = \frac{A_l}{\log_{2}\left(M_l\right)R_lR_s}.
\end{equation}
Consequently, the total transmission delay can be evaluated as
\begin{equation}
T = \sum_{l=1}^L T_l = \sum_{l=1}^L \frac{A_l}{\log_{2}\left(M_l\right)R_lR_s}. 
\end{equation}

Let $\mathcal{M}_l$ be the modulation scheme chosen from a modulation candidate $\mathcal{M}_{all}$ for the transmission of the $l$th frame. Let $\mathcal{C}_l$ be the coding scheme chosen from a coding candidate $\mathcal{C}_{all}$ for the transmission of the $l$th frame. 
The modulation scheme $M_l$ will determine the symbol error rate (i.e., $P_{l,sym}\left(\mathcal{M}_l\right)$) and the number of bits per symbol (i.e., $\log(M_l)$). The coding scheme $C_l$ will determine error-correcting capabilities (i.e., $C_l$) and code rates (i.e., $R_l$). Both of them are key factors that dominate the semantic loss and transmission delay. Aiming at minimizing the semantic loss within a transmission delay constraint, we formulate the semantic-aware adaptive modulation and coding problem as
 \begin{equation}\label{eq11}
 \begin{split}
 \min&~SL\\
 s.t.: ~&T\leq T_{th},\\
      &\mathcal{M}_l\in \mathcal{M}_{all},\\
      &\mathcal{C}_l \in\mathcal{C}_{all}.
 \end{split}
 \end{equation}
 
The problem delineated in (\ref{eq11}) is inherently a discrete optimization problem. To determine the optimal solution, one could theoretically conduct an exhaustive search of all possible solutions. This approach, however, escalates in computational complexity exponentially with the number of frames, following the formula  $\mathcal{O}(|\mathcal{M}_{all}|^L|\mathcal{C}_{all}|^LC_1)$, where  $C_1$ signifies the complexity of computing
$SL$ and $T$ for each modulation and coding combination.

In order to alleviate the computational burden of this method, we propose a semantic-aware greedy search method. 
We randomly generate many feasible modulation and coding candidates and output the one with minimum semantic loss and cceptable transmission delay below the threshold. By applying such greedy search method, the computational complexity is significantly reduced to 
 $\mathcal{O}
\left(N_{set}LC_1\right)$, where $N_{set}$ is the number of candidates the algorithm searches. This method provides a more computationally efficient approach, striking a balance between achieving near-optimal solutions and adhering to constraints of semantic importance and transmission delay.

\section{Semantic Correction and Visual Data Generation}
The process of receiving semantically important signal representations, which are detected as bits, involves error correction for rectifiable errors, while uncorrectable frames are discarded. This study proposes a novel approach that leverages a LLM for word completion in scenarios where frames are lost.

\subsection{Semantic Correction Methodology}
Despite efforts to prioritize the transmission of semantically significant frames and quantify their importance, the reliability of information transmission cannot be guaranteed due to unpredictable factors such as wireless channel fading and noise interference. These factors can distort or result in the loss of information during transmission. In the ULSC process, the intermediate state is the natural language description. If this is not transmitted reliably, it can hinder the receiver's understanding and decision-making. This issue is particularly critical in applications involving visually impaired individuals, where misunderstandings or delays could compromise personal safety. Therefore, addressing the challenge of completing missing frames in received natural language descriptions is crucial for ensuring reliable semantic communication.

To tackle this challenge, this study proposes the use of a pre-trained LLM, such as BERT, which leverages contextual information and pre-training knowledge to infer and recover missing frames. 
The specific steps of this process are outlined as follows.
First, the input sentences are preprocessed to fit the BERT model. Suppose we have a sentence: ``A beach with palm [MASK] and clear blue water.", where [MASK] indicates the missing word. Using BERT's tokenizer, the sentence is tokenized and special tokens [CLS] and [SEP] are added to denote the start and end of the sentence, respectively. Next, the segmented sentence is converted into the corresponding ID in the vocabulary, which is used as input to the model. In addition, to accommodate the input format of BERT, these IDs need to be converted into tensor form. The preprocessed input data is fed into the BERT model. The BERT model consists of multiple Transformer layers, each of which is processed by Self-Attention mechanisms and Feed Forward Network. Among them, the mathematical expression of the Self-Attention mechanism is:
\begin{equation}\label{ATTEN}
{\rm Attention}(Q,K,V) = {\rm softmax} \left( \frac{QK^T}{\sqrt{d_k}}\right)V.
\end{equation}
$Q$, $K$ and $V$ are query, key and value matrices respectively, and $d_k$ is the dimension of the key vector. In BERT, each word representation interacts with other word representations through this mechanism to capture global information in the sentence. After passing the last layer of BERT, a contextual representation of each word can be obtained. Suppose the output is represented as a matrix $\boldsymbol{\Pi}$ with the shape ($L_\Pi$, $D_\Pi$), where $L_\Pi$ is the length of the sentence and $D_\Pi$ is the dimension of the representation. It can be expressed as:
\begin{equation}\label{mp}
{\boldsymbol{\Pi}} = \left[\boldsymbol{\pi}_{[{\rm CLS}]},\boldsymbol{\pi}_{[{\rm A}]},\cdots,\boldsymbol{\pi}_{[{\rm MASK}]},\cdots, \boldsymbol{\pi}_{[{\rm water}]},\boldsymbol{\pi}_{[{\rm SEP}]}\right].
\end{equation}
Next, the representation at position is extracted, and it is transformed into a probability distribution over the vocabulary through a fully connected layer and a softmax layer:
\begin{equation}\label{pd}
{\rm \textbf{P}}(w|{\rm context}) = {\rm softmax}(\boldsymbol{\pi}_{[{\rm MASK}]}W+b),
\end{equation}
where $W$ and $b$ are trainable weights and bias parameters, and ${\rm \textbf{P}}(w|{\rm context})$ is the probability distribution of each word in the vocabulary. Finally, the word corresponding to the maximum value in the probability distribution is selected as the prediction result:
\begin{equation}\label{rs}
\hat{w} = {\rm argmax}_w {\rm \textbf{P}}(w|{\rm context}).
\end{equation}
By aligning the correction mechanism with human cognitive processes, this approach ensures a more reliable transmission of semantic information, thereby enhancing the overall effectiveness of communication systems, especially in critical applications.

\begin{figure}
\centering  
\includegraphics[width=0.95\linewidth]{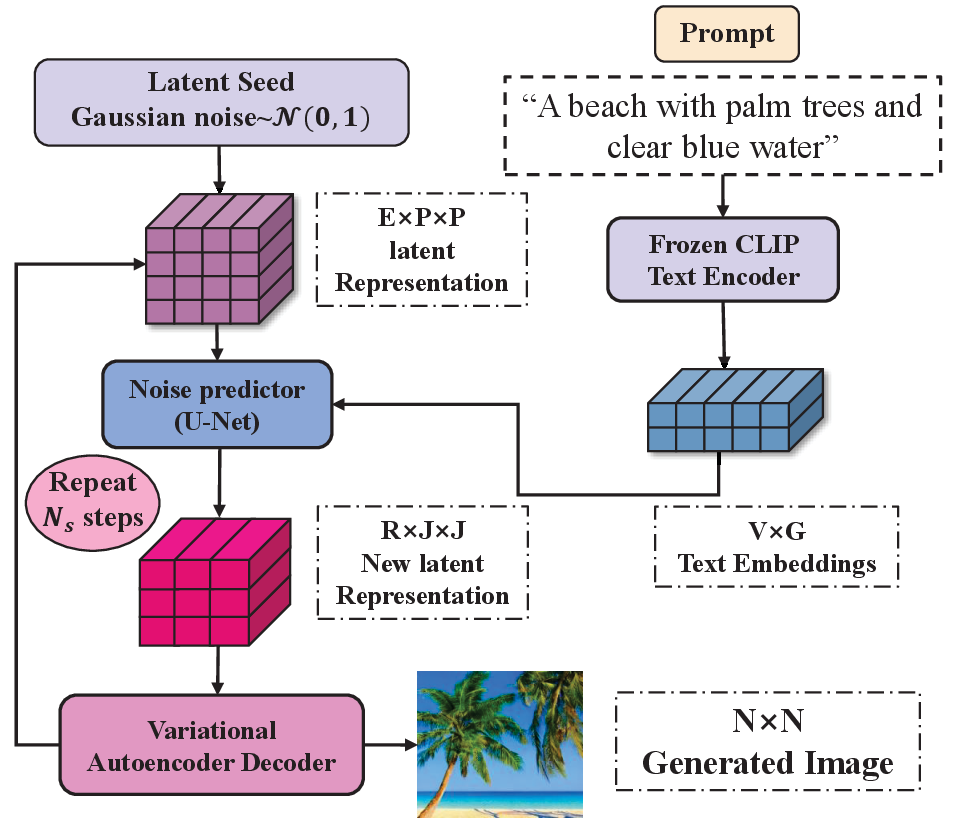}
\caption{Visual data generation.}
\label{t2i}
\end{figure}

\subsection{Visual Data Generation}
Visual data generation is a pivotal component in enabling smart agents to interpret semantics and contextual nuances in received text. This process is achieved through sophisticated neural network models that transform textual descriptions into corresponding visual representations, such as images. Achieving a harmonious and logical alignment between the generated images and their textual counterparts is a challenging endeavor due to the inherent ambiguity in correlating images with natural language expressions and the necessity to maintain a high level of image realism.

Prominent developments in this field include OpenAI's DALL-E model\cite{ramesh2021zero} and Google's Imagen model\cite{saharia2022photorealistic}, both designed for text-to-image generation. A notable limitation in these models is their computational efficiency, primarily because they function in pixel space. To manage computational constraints, these models initially produce images of $64\times64$ resolution, which are subsequently upscaled to $256\times256$ and finally to $1024\times1024$, utilizing super-resolution techniques. A significant advancement is presented in \cite{rombach2022high}, which introduced the Stable Diffusion model. The core idea is to start from the pure noise image and gradually recover to the target image through a series of reverse diffusion steps. Its operation in latent space offers enhanced computational efficiency, making it a superior choice for integration into the ULSC system.

The image generation process of the Stable Diffusion model is illustrated in Fig.\ref{t2i}. First, the user inputs a text prompt, such as ``a beach with palm trees and clear blue water". This text prompt is processed through a pre-trained CLIP text encoder, generating the corresponding text embeddings. The CLIP text encoder converts the text into a multi-dimensional vector that effectively captures the semantic information of the text. Meanwhile, an initial noise seed is sampled from a standard Gaussian distribution. These noise seeds serve as the initial input for the latent representation. The noise seeds are represented as a multi-dimensional matrix of a specific size, forming the starting point of the image generation process. Next, the noise seeds and text embeddings are fed into the noise predictor, which uses a U-Net model\cite{ronneberger2015u}. The U-Net model is a convolutional neural network capable of denoising while preserving crucial details in the image. At each timestep, the U-Net model predicts and removes noise based on the current latent representation and text embeddings, gradually generating a new latent representation. This process involves multiple iterative steps, with each step progressively removing some noise, transforming the latent representation from pure noise to a form closer to the target image. Throughout the denoising process, the text embeddings provide guidance for image generation, ensuring that the generated image aligns with the input text prompt. After completing all the denoising steps, the final latent representation is decoded through a VAE decoder, producing the final high-quality image. The VAE decoder converts the denoised latent representation into a complete image, significantly enhancing the detail and quality of the image, which matches the initial text prompt. Through this process, the Stable Diffusion model can start from random noise, incorporate the input text prompt, and gradually generate a high-quality image that corresponds to the description. This process fully leverages the powerful features of pre-trained models, particularly the CLIP text encoder and the U-Net denoising network, ultimately achieving an efficient transformation from text to image.

\section{Experiments and Discussions}
In this section, we will demonstrate the advantage of semantic importance-aware communications through experiments and that of using LLM to assist semantic error correction. In particular, adaptive modulation and coding that is compatible with existing communication systems. Besides, we also demonstrate the advantage of ULSC compared to feature-based semantic communication in achieving high semantic similarity and protecting privacy.

\subsection{Experimental Setup}
\subsubsection{Datasets}
We collected a comprehensive dataset comprising $100$ images sourced from the Internet, aimed at validating the effectiveness of our newly proposed method. This dataset comprises landscapes, animals, people, cartoons, and various other image categories. The resolution of each image is $256\times 256$. Additionally, in our experiments,  we select a deep learning-based algorithm for comparison. This algorithm requires extensive training on large-scale dataset. To achieve excellent algorithm performance, we train the algorithm using the ImageNet dataset\cite{deng2009imagenet}. ImageNet is a large-scale image dataset covering over $1,000$ categories, ranging from animals to household items and various other categories. These images are not only diverse but also include objects from different angles and in various environments, making it an ideal dataset for training deep learning models. In our experiments, we adjust the resolution of each ImageNet image to $256\times 256$.

\subsubsection{Baselines}
We compare our proposed method with DeepJSCC\cite{9953099}, which shows excellent performance in wireless image transmission. In DeepJSCC, convolutional neural networks (CNNs) are employed to encode the source image into a compact representation that is directly transmitted over a communication channel. The receiver uses another CNNs to decode the received signal back into the original image. In our experiments, we use a frame erase channel to simulate real-world wireless communication conditions. For the features transmitted in DeepJSCC, we investigate two packaging schemes: one where all features were packaged into $7$ frames (denoted as ``DeepJSCC with $7$ frames''), and another where they were packaged into $14$ frames (denoted as ``DeepJSCC with $14$ frames''). This is because the natural language generated from images in the proposed method contains a maximum of $14$ words and a minimum of $7$ words.

\begin{figure}
\centering  
\includegraphics[width=0.95\linewidth]{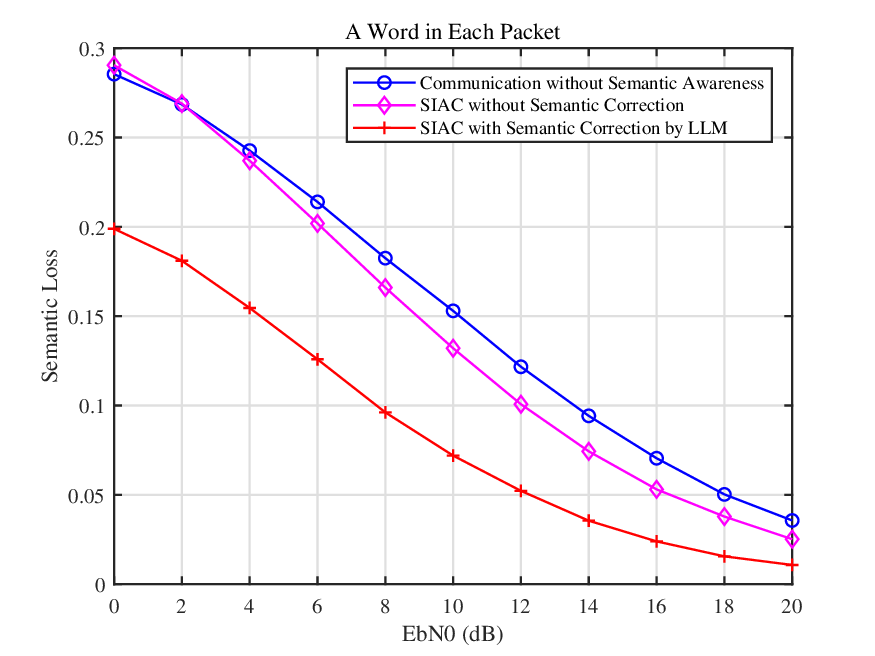}
\caption{Semantic loss of different transmission strategies.}
\label{results1}
\end{figure}

We further validate the privacy protection capabilities of our proposed method using model inversion attacks. Model inversion attacks aim to reconstruct the original input data from the output of the model, potentially revealing sensitive visual information. In our evaluation, we apply model inversion attacks to both our proposed method and the comparison method, DeepJSCC. By analyzing the reconstructed images, we can determine how much information can be inferred about the original input.

The steps to implement a model inversion attack on DeepJSCC are as follows. First, the output data is collected, which in the case of DeepJSCC are the feature representations output by the encoder. Then, CNNs are designed as an inversion model, aiming to map the output data back to the original input data. Next, the inversion model is trained based on the collected data. During training, the inversion model learns to reconstruct the original input from the output data. Finally, the trained inversion model is used to perform the inversion attack on the output data from DeepJSCC, generating reconstructed input data. For the inversion attack of our proposed method, we assume that after stealing the signals, the same generative model is used to generate visual data.

\subsubsection{Metrics}
In this paper, we focus on semantic similarities between transmitted and recovered semantic understanding. The commonly used peak signal-to-noise ratio (PSNR) mainly measures the difference of images at the pixel level, focusing on the average error between the original image and the reconstructed image. An image that scores high on the PSNR may not necessarily be visually closer to the original image. This is mentioned ``semantic gap", namely the image pixel level is similar, but there were significant differences in content and semantics. In this paper, we propose to calculate the image similarity between the original image and the generated image to evaluate the performance of the algorithm.
\begin{equation}\label{ss}
{\rm Image \, Similarity}(\mathbf{S}, \hat{\mathbf{S}}) = \frac{\textbf{\emph{C}}_\Phi(\mathbf{S}) \cdot \textbf{\emph{C}}_\Phi(\hat{\mathbf{S}})} {\left \| \textbf{\emph{C}}_\Phi(\mathbf{S}) \right \|  \| \textbf{\emph{C}}_\Phi (\hat{\mathbf{S}})  \|},
\end{equation}
where $\textbf{\emph{C}}_\Phi$, representing CLIP, is a huge pre-trained model
including billions of parameters for multimodal image and text embedding.
The image similarity defined in \eqref{ss} is a value between $0$ and $1$, indicating the degree of semantic similarity between the generated image and the original image, where $1$ indicates the highest similarity and $0$ indicates no similarity.

In addition, the purpose of the model inversion attack is to reconstruct the original image from the signal, thus violating the user's visual privacy. Therefore, we analyze the quality of the image reconstructed by the hypothetical adversary as a measure of the privacy protection ability of the algorithm. Specifically, we use PSNR, which is expressed as
\begin{equation}\label{psnr}
{\rm Attacker\text{-}PSNR} = 10 \log_{10} \frac{\rm MAX^2}{\rm MSE} \quad {\rm (dB)},
\end{equation}
where ${\rm MSE} = d(\mathbf{S}, \hat{\mathbf{S}})$ is the mean squared-error between the original image $\mathbf{S}$ and the generated image $\hat{\mathbf{S}}$, and ${\rm MAX}$ is the maximum value of the image pixels. The images we used are $24$-bit depth RGB images with each color channel having $8$ bits per pixel, so ${\rm MAX} = 2^8-1 = 255$. The lower the PSNR, the better the privacy protection.

\begin{figure}
\centering  
\includegraphics[width=0.95\linewidth]{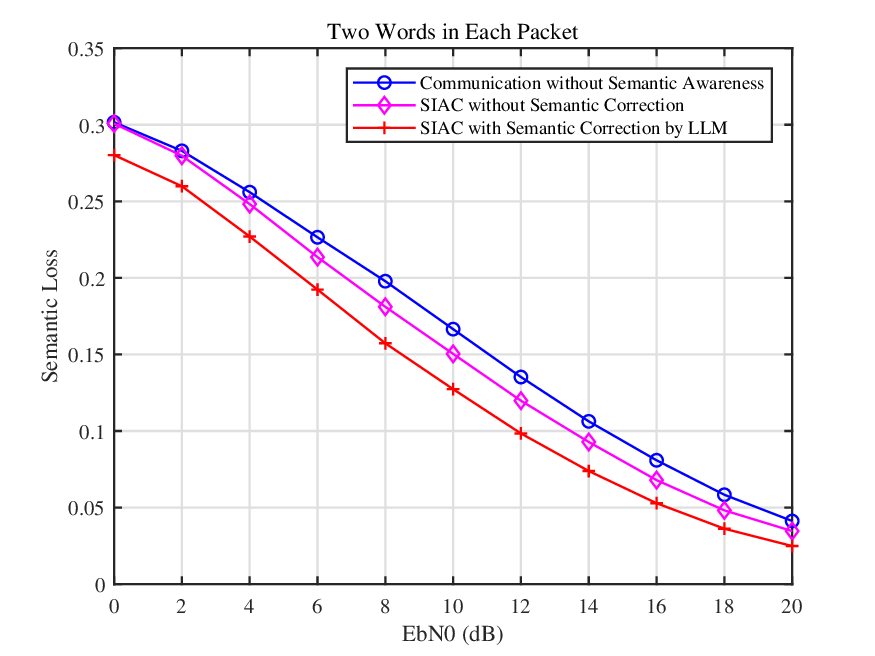}
\caption{Semantic loss of different transmission strategies.}
\label{results2}
\end{figure}

\subsection{Superiority of Semantic-Aware Adaptive Modulation and Coding}

The modulation scheme is selected from the set $\mathcal{M}_{all}=\{\rm{BPSK}, \rm{QPSK}, \rm{16QAM}, \rm{64QAM}, \rm{256QAM}\}$. The channel coding schemes is chosen from $\mathcal{C}_{all}=\{(2^{k}, k,2^{k-1})
\textrm{~Hadamard~codes}|k=3,4,5,6\}$ with block length $B_l=2^{k}$, code rate $R_l=k/2^k$ and minimum distance $d_{l,\min}=2^{k-1}$. It is worth noting that Hadamard codes are chosen as it has closed-form expressions of block length, coding rate, and minimum codeword distance. The proposed AMC can also be extended to selecting other MCS. The frame header is assumed to be $10$ Bytes.  In the experiment, we investigate two packeting solutions, one packing one word in a packet, and the other packing two words in a packet. The delay threshold is set to $1$ ms and $2$ ms for transmitting all packets for the above two cases, respectively.  The MCS candidate set size $N_{set}$ is set to $1000$. The experiment results are illustrated in Figs. \ref{results1} and  \ref{results2}, respectively. It is showed that in both experiment setups, the semantic-aware communication scheme considerably outperforms conventional communication without semantic awareness. With the assistance of LLM at the receiver for semantic error correction, the semantic loss can be greatly reduced, reflecting the effectiveness of LLM for communication performance improvement. The performance gain in a word in each packet is more significant as in this case, the LLM can handle one word missing compared to two words missing.

\begin{figure}
\centering  
\includegraphics[width=0.95\linewidth]{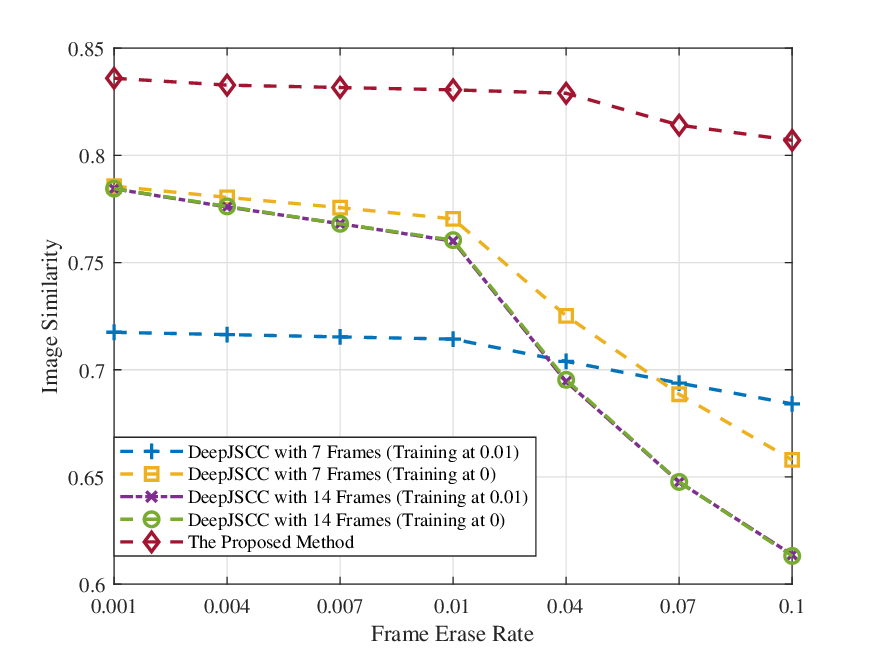}
\caption{Image similarity of different methods versus frame erase rate.}
\label{results3}
\end{figure}

\subsection{Comparison with Feature Level Semantic Communications}

In this subsection, we evaluate the performance of the proposed method and the baseline methods at different frame erase rates, that is, semantic similarities between transmitted and recovered semantic understanding. If not specified otherwise, the frame erase rates during testing are $0.001$, $0.004$, $0.007$, $0.01$,  $0.04$,  $0.07$, and $0.1$. For the baseline method \emph{DeepJSCC with 7 Frames}, We assume that it is trained at the frame erase rate of $0$ and $0.01$, respectively. The same is true for \emph{DeepJSCC with 14 Frames}.

Fig.\ref{results3} shows the image similarity of different methods versus frame erase rate. It can be seen from that there are significant differences in image similarity performance between different methods under different frame erase rates. When the frame erase rate is low (e.g. $0.001$ to $0.01$), the image similarity of \emph{DeepJSCC with 7 Frames (Training at 0.01)} remains about $0.72$.  With the increase of frame erase rate, the image similarity decreases slightly, but the whole image remains relatively stable. In contrast, at low frame erase rates (e.g. $0.001$ to $0.01$), the image similarity of \emph{DeepJSCC with 7 Frames (Training at 0)} is around $0.78$, but when frame erase rates reach $0.1$, the image similarity drops significantly to around $0.65$, showing sensitivity to high frame erase rates. Compared to \emph{DeepJSCC with 7 Frames}, \emph{DeepJSCC with 14 Frames} performs slightly better in most cases. However, with the increase of frame erase rate, its image similarity decreases rapidly, especially under the frame erase rate of $0.1$. In contrast, the proposed method exhibits the best image similarity at all frame erase rates. On a scale of $0.001$ to $0.1$, image similarity remains between $0.82$ and $0.84$. Even at a high frame erase rate of $0.1$, the image similarity of this method is much higher than that of other methods. On the whole, the proposed method is significantly better than other methods in each frame erase rate, showing better robustness and stability.

\subsection{Advantages in Protecting Privacy}

\begin{figure}
\centering  
\includegraphics[width=0.95\linewidth]{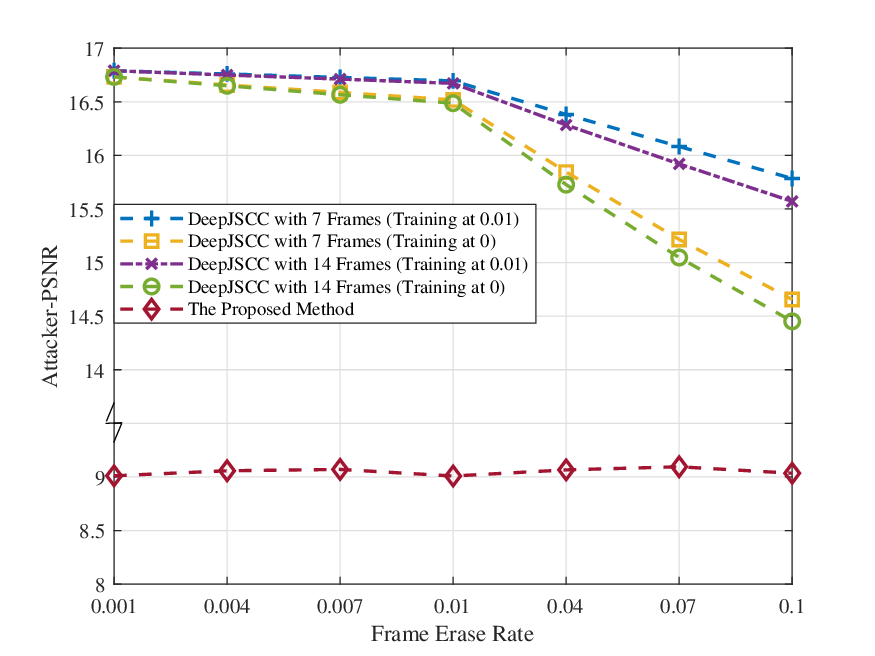}
\caption{Attacker-PSNR of different methods versus frame erase rate.}
\label{results4}
\end{figure}

To verify the advantages of the proposed method in protecting privacy, we conduct model inversion attacks on the proposed method and the baseline methods. In Fig.\ref{results4}, we show the Attacker-PSNR versus frame erase rate. We can observe that different methods have significant differences in Attacker-PSNR under different frame erase rates. For \emph{DeepJSCC with 7 Frames} and \emph{DeepJSCC with 14 Frames} trained at a frame erase rate of $0.01$, the Attacker-PSNR remains around $16.8$ at low frame erase rates (e.g., $0.001$ to $0.01$) with little variation. As the frame erase rate increases, the Attacker-PSNR slightly decreases but generally stays around $16$. For \emph{DeepJSCC with 7 Frames} and \emph{DeepJSCC with 14 Frames} trained at a frame erase rate of $0$, the Attacker-PSNR decreases more rapidly with increasing frame erase rate, especially dropping to around $15$ at a frame erase rate of $0.1$. In contrast, the proposed method has significantly lower Attacker-PSNR at all frame erase rates than the baseline methods, with Attacker-PSNR remaining between $8.5$ and $9$ in the range of $0.001$ to $0.1$. This shows that the proposed method makes it difficult for attackers to obtain high-quality images regardless of how frame erase rates vary. 

To more intuitively demonstrate the privacy benefits of our proposed approach, we present images obtained by adversaries using model inversion. We can see that the image obtained by attacking the baseline method is the same person as the original image, although there is some blur and noise.  Although the image obtained by attacking the proposed method is clear, it can be clearly seen that it is different from the person in the original image.  This effectively protects the user's visual privacy, indicating the advantages of the proposed method.


\section{Conclusion}
In this study, we introduced an innovative approach for understanding-level semantic communications by integrating semantic importance-aware communication (SIAC) with semantic correction using advanced LLMs such as BERT and ChatGPT. Our method aligns image communication within the human-comprehensible semantic space of natural language, enabling effective semantic correction and significance analysis through LLMs. We developed a technique to quantify the semantic value of data in each frame, supporting a semantically informed SIAC strategy. Additionally, the research explores a semantic-aware adaptive modulation and coding scheme, complemented by a proposed low-complexity greedy search method to ensure semantically robust communications within set latency limits. The utilization of LLMs for filling in missing frames is also introduced. Experimental results validate the superiority of our approach in enhancing agent-human communications or generative task-oriented communications, particularly in terms of minimizing semantic loss and bolstering privacy protection.

\begin{figure*}
       \centering       \includegraphics[width=1\linewidth]{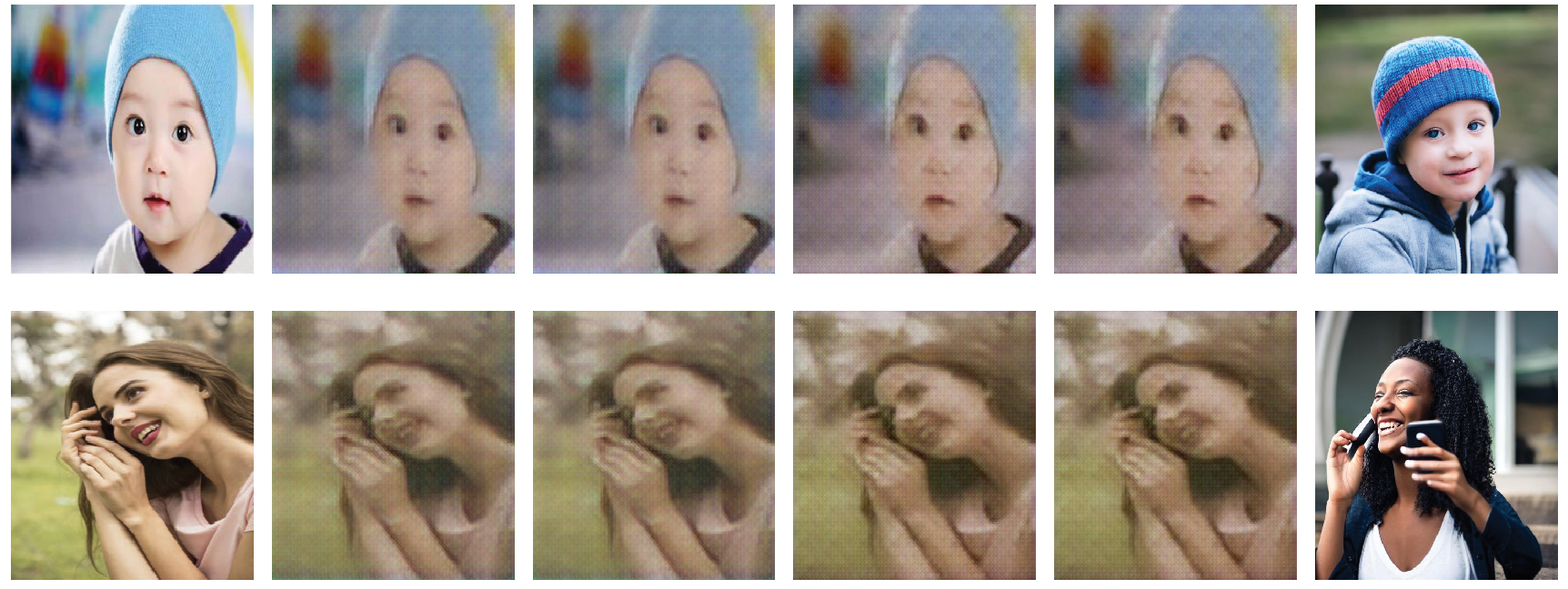}       \caption{
From left to right are the original image, and the images obtained by model inversion attacks on \emph{DeepJSCC with 7 Frames (Training at 0.01)}, \emph{DeepJSCC with 7 Frames (Training at 0)}, \emph{DeepJSCC with 14 Frames (Training at 0.01)}, \emph{DeepJSCC with 14 Frames (Training at 0)}, and the proposed method. For each reconstructed image in evaluation, the frame erase rate is set to $0.001$.}
       \label{result5}
\end{figure*}

\bibliographystyle{IEEEtran} 
\bibliography{IEEEabrv,bib}

\end{document}